\documentclass[aps,pra,reprint,superscriptaddress, longbibliography]{revtex4-2}
\usepackage{lipsum, babel}

\usepackage{siunitx}
\usepackage{amsmath,bm}
\usepackage{amsbsy}
\usepackage{amssymb}
\usepackage{mathptmx, textcomp}
\usepackage{color}
\usepackage{braket}
\usepackage{graphicx}
\usepackage{textcomp}
\usepackage{notes2bib}
\usepackage{textcomp}
\usepackage{mwe}
\usepackage{ulem}
\usepackage{siunitx}
\usepackage{filecontents}
\usepackage{soul}

\definecolor{bl}{rgb}{0, .1, .6}
\usepackage[colorlinks=true, citecolor = bl, linkcolor = bl, urlcolor=bl, pdfborder={0 0 0}]{hyperref}

\DeclareSIUnit\gauss{G}

\usepackage{xcolor}

\begin{document}
\title{Laser driven superradiant ensembles of two-level atoms near Dicke's regime}

\author{G. Ferioli}
\email{giovanni.ferioli@institutoptique.fr}
\affiliation{Universit\'e Paris-Saclay, Institut d'Optique Graduate School, CNRS, 
Laboratoire Charles Fabry, 91127, Palaiseau, France}
\author{A. Glicenstein}
\affiliation{Universit\'e Paris-Saclay, Institut d'Optique Graduate School, CNRS, 
Laboratoire Charles Fabry, 91127, Palaiseau, France}
\author{F. Robicheaux}
\email{robichf@purdue.edu}
\affiliation{Department of Physics and Astronomy, Purdue University,
West Lafayette, IN 47907, USA}
\affiliation{Purdue Quantum Science and Engineering Institute, Purdue
University, West Lafayette, IN 47907, USA}
\author{R.~T.~Sutherland}
\email{robert.sutherland@utsa.edu}
\affiliation{Department of Electrical and Computer Engineering, Department of Physics and Astronomy, University of Texas at San Antonio, San Antonio, TX 78249, USA}
\author{A. Browaeys}
\affiliation{Universit\'e Paris-Saclay, Institut d'Optique Graduate School, CNRS, 
Laboratoire Charles Fabry, 91127, Palaiseau, France}
\author{I. Ferrier-Barbut}
\affiliation{Universit\'e Paris-Saclay, Institut d'Optique Graduate School, CNRS, 
Laboratoire Charles Fabry, 91127, Palaiseau, France}
\begin{abstract}
We report the experimental observation of superradiant emission emanating from an elongated dense ensemble of laser cooled two-level atoms, with a radial extent smaller than the transition wavelength.
In the presence of a strong driving laser, we observe that the system is superradiant along its symmmetry axis. This occurs even though the driving laser is orthogonal to the superradiance direction. This superradiance modifies the spontaneous emission, and, resultantly, the Rabi oscillations. We also investigate Dicke superradiance in the emission of an almost fully-inverted system as a function of atom numnber. The experimental results are in qualitative agreement with ab-initio, beyond-mean-field calculations.  
\end{abstract}

\maketitle
\normalem
In 1954, Dicke predicted that the radiation emitted by a dense ensemble of atoms should be dramatically different than the emission from independent atoms \cite{dicke1954}. According to Dicke, the decay of a fully inverted cloud of $N$ emitters confined in a region smaller than their transition wavelength is characterized by a burst of radiation with peak intensity scaling $\propto N^{2}$, rather than the expected $\propto N$. This behavior, known as superradiance (or superfluorescence), has been investigated in many experimental platforms including low density clouds of atoms or molecules \cite{gross1982,skribanowitz1973observation, gross1976observation, gibbs1977single,araujo2016superradiance, roof2016observation,das2020subradiance}, semiconductors \cite{scheibner2007superradiance,cong2016Dicke}, nuclei \cite{rohlsberger2010collective}, superconducting qubits \cite{mlynek2014observation} and Rydberg gases \cite{gross1979maser, wang2007superradiance, day2008dynamics, grimes2017direct}. Recently, interest in superradiance has grown, following theoretical proposals \cite{meiser2009prospects,maier2014a} and experiments \cite{bohnet2012steady, norcia2016cold, norcia2016superradiance, norcia2018frequency,laske2019,schaffer2020} that describe how superradiance could help realize a novel class of ultra-stable lasers. \par

The study of superradiant effects\textemdash with an external driving field\textemdash constitutes a new direction of research that extends beyond Dicke's original proposal. 
In the presence of driving, the cloud of emitters can be mapped onto a driven-dissipative spin system where the interplay between dissipation, driving, and collective effects could lead to novel many-body quantum phases \cite{weimer2015variational, olmos2014steady, parmee2018phases, lee2013unconventional}. Motivated by this, we here investigate 
the coherent emission of a dense, elongated and  microscopic cloud of (effectively) two-level $^{87}$Rb atoms in the presence of an on-resonance external drive.

In our setup, atoms are trapped in a cylindrically-symmetric volume, with radial dimension smaller than the transition's resonant wavelength. This modifies spontaneous emission \textit{in the axial direction} of the cloud, with all $N$ atoms emitting collectively along this direction \cite{gross1982}, while emission \textit{in the radial directions} is not collective. This strong axial coupling creates a situation akin to that of an atomic cloud coupled to the mode of an optical cavity \cite{black2003collective, nagorny2003,Baumann2010Dicke,ritsch2013cold}. We demonstrate that this system undergoes Rabi oscillations that are modified by superradiance, where the amount of light scattered along the axis of the cloud is enhanced, although driving is performed perpendicularly to the axis. We compare our experimental results with ab-initio numerical simulations based on a second-order cumulant expansion technique \cite{robicheaux2021lowest, plankensteiner2021quantumcumulantsjl}, finding qualitative agreement. Finally, through tuning the duration of the driving field, we achieve almost full inversion. This allows us to study the subsequent decay, observing features typical of Dicke superradiance in this dense regime where the influence of the resonant dipole-dipole interactions between atoms remains under debate \cite{gross1982, FRIEDBERG1972365,zhou_2016, superradiance2017sutherland}.

\begin{figure*}[!]
\minipage{0.23\textwidth}
\includegraphics[width=0.95\linewidth]{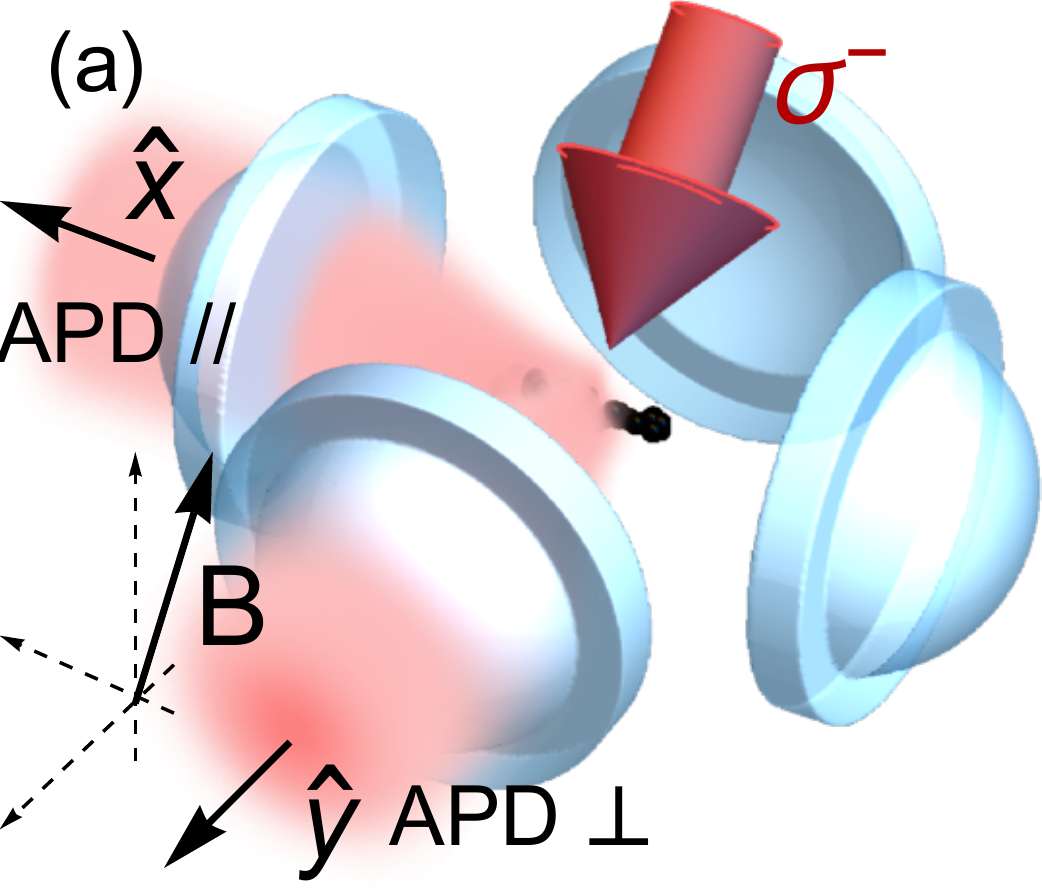}
\endminipage\hfill
\minipage{0.75\textwidth}
\includegraphics[width=\linewidth]{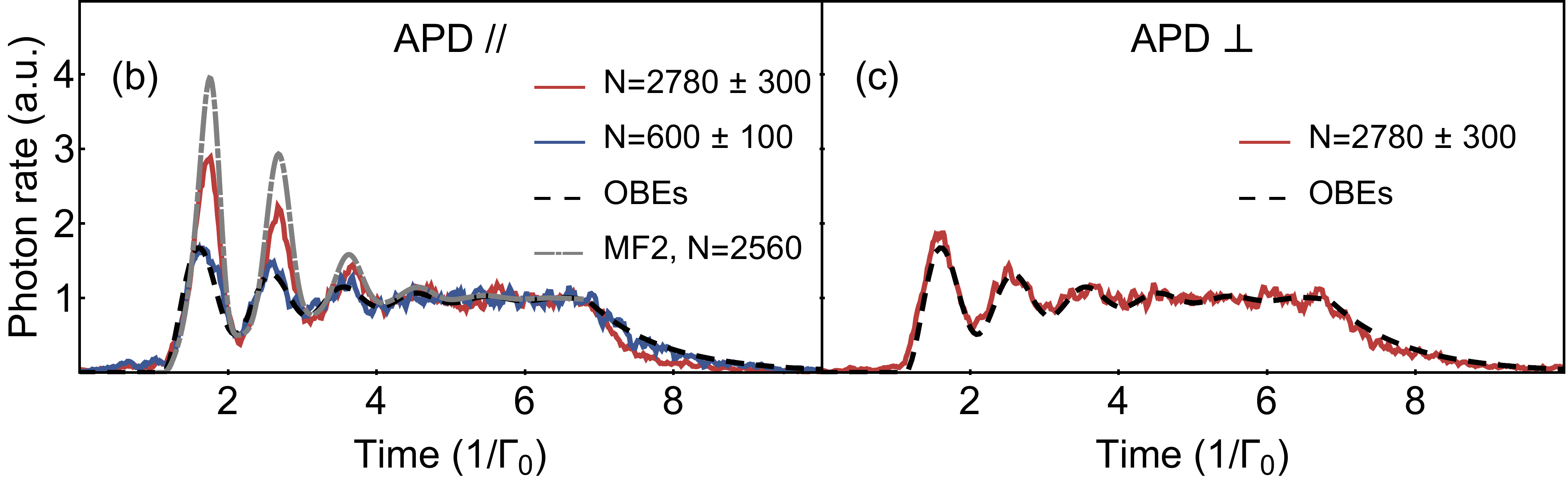}
\endminipage\hfill
	\caption{{\bf Experimental setup and observation of collective Rabi oscillations}.
	(a) Sketch of the experimental setup. The excitation beam is aligned along the magnetic field
	$\bf B$ and propagates along $\hat{z}-\hat{y}$, perpendicularly to the main axis of the cloud. The light emitted by the atoms is collected by two APDs, one along the main axis of the cloud ($\hat{x}$), the other one perpendicular to it ($\hat{y}$). 
	(b) Photon rate along the axis of the cloud versus time. For low $N$ (blue solid line) the dynamics is reproduced by the solution of the optical Bloch equations for a single atom (black dashed). For large $N$ (red solid), the experimental results agree qualitatively with MF2 calculations (grey dot-dashed).
	(c) Photon rates measured in the radial direction for $N=2780$ (red solid). In this direction, the dynamics remains consistent with the single-atom optical Bloch equations (black dashed) for all $N$.}
	\label{fig1}
\end{figure*}

Our experimental setup, detailed in \cite{setup, Glicenstein2020, ferioli2020storage}, relies on four high-numerical-aperture (NA) aspherical lenses, as sketched in Fig.\,\ref{fig1}(a). We load up to $5000$ $^{87}$Rb atoms in a $\SI{2.5}{\micro\meter}$ waist, \SI{7.5}{\milli\kelvin}-deep optical trap. The atomic cloud has an approximate temperature of $\SI{650}{\micro \kelvin}$, a $1/e^2$-radial size estimated to be $\ell_{\text{rad}}\simeq 0.5\, \lambda_0$ and an axial size measured to be $\ell_{\text{ax}}\simeq 15 \,\lambda_0$ \footnote{The measured trapping frequencies are $\omega_r\simeq 2\pi\times \SI{100}{\kilo\hertz}$ and $\omega_ax\simeq 2\pi\times \SI{10}{\kilo\hertz}$ which should lead to a density distribution with an axial size smaller than the measured one by a factor 3. We assign this disagreement to some imperfections of the trapping beam.}. By applying an external magnetic field of $\SI{50}{\gauss}$ and performing hyperfine and Zeeman optical pumping with the same polarization as the excitation light, we isolate a closed transition between the states $\ket{g}=\ket{5S_{1/2},F=2,m_{\text{F}}=-2}$ and $\ket{e}=\ket{5P_{3/2},F'=3,m_{\text{F}}'=-3}$, forming a cloud of two-level emitters. The system is excited perpendicularly to the main axis of the cloud using $\sigma^-$ polarized light resonant with the $D_2$ transition of $^{87}$Rb (
$\lambda_0\simeq \SI{780}{\nano\meter}$, $\Gamma_0\simeq 2\pi \times \SI{6.1}{\mega\hertz}$ and $I_{\text{sat}}\simeq \SI{1.67}{\milli\watt/\centi\meter^2} $). Since the excitation beam is much larger than the cloud, all atoms experience approximately the same light intensity. We collect the fluorescence emitted by the cloud into two fiber-coupled avalanche photodiodes (APD) in single-photon counting mode, one aligned along the axial direction of the cloud ($\hat{x}$ direction of Fig.\ref{fig1}(a), APD //) and the other perpendicularly to it ($\hat{y}$ direction of Fig.\ref{fig1}(a), APD $\perp$). The photon rates reported in this work represent the number of photons collected by the APDs in $\SI{1}{\nano\second}$ time bins. The temporal profile of the excitation beam is shaped by means of a fiber electro-optic modulator (EOM) permitting a fast switching-off time, shorter than $\SI{1}{\nano\second}$ \footnote{The typical extinction ratio obtained with the EOM is $1\%$, the remaining light is turned off by means of two acousto-optic modulators in series, in a timescale of $\SI{10}{\nano\second}$}. We apply the same excitation pulse 20 times on the same cloud, checking that the fraction of atoms lost during the process is less than $10\%$. To achieve a sufficiently high signal, we repeat this sequence on 1500 to 3000 different clouds, loaded at a rate of $\SI{2}{\hertz}$.\par

We first investigate the influence of superradiance on Rabi oscillations. 
We excite the cloud (in free space) with a pulse of duration $\SI{150}{\nano\second}\simeq6/\Gamma_0$, 
sufficiently long to reach steady-state. In this work, the excitation beam has a saturation parameter $s=I/I_{\text{sat}}\simeq 85$. We collect the emitted photons both in the axial and radial directions. Examples of the recorded photon rates, normalized to the steady-state values, are reported in Fig.\,\ref{fig1}(b,c) \footnote{We checked that the steady-state value increases linearly with $N$ in the low-atom number regime, as expected}. \par 

\begin{figure}[!]
	\centerline{\includegraphics[width=0.95\columnwidth]{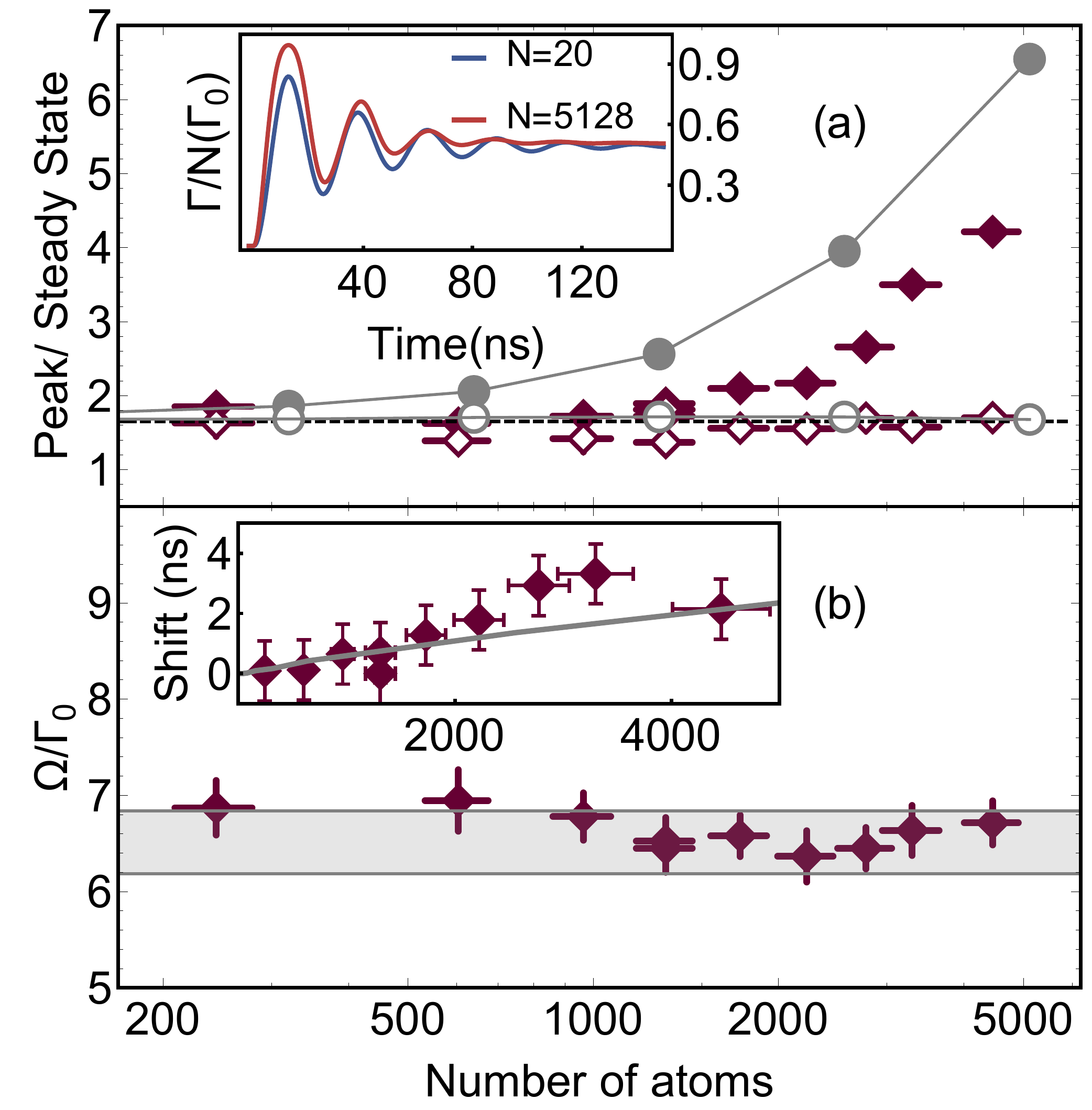}}
	\caption{{\bf Observation of collective Rabi oscillations}. (a) Filled (empty) diamonds: measured ratios of the peak to steady-state emission rates for the collective Rabi oscillations recorded along the axial (radial) direction of the cloud. 
	Gray points: results of the numerical simulations performed with the MF2 model (see text). The vertical error bars represent the standard error in the estimation of the steady state (smaller than symbols).
	Black dashed line: results from the OBEs. 
	Inset: total photon emission rate (in a $4\pi$ solid angle) per atom $\Gamma(t)/N$, calculated with MF2, for small and 
	large  $N$. 
	(b) Diamonds: measured Rabi frequencies. The error bars represent the variance of the Gaussian distribution used to fit the experimental spectra. Gray area: expected value for the single atom Rabi frequency $\Omega/\Gamma_0=\sqrt{s/2}$, 
	including the experimental error on the intensity of the excitation beam. Inset: delay of the position of the maximum at the first Rabi fringe versus atom number (gray line, MF2 simulations). Error bars show the finite time resolution of the detector ($\SI{1}{\nano\second}$).}
	\label{fig2}
\end{figure}

In the low $N$ regime, we observe that the cloud behaves as an ensemble of non-interacting emitters. Indeed, the dynamics of the system is well described by the single-atom optical Bloch equations (OBEs), as can be seen in Fig.~\ref{fig1}(b). In the {\it axial} direction, as $N$ increases, the interplay between superradiance and driving by the laser enhances the observed emission peaks during Rabi oscillations (colored filled diamonds in Fig.\,\ref{fig2}(a)). Interestingly, this effect is absent in the {\it radial} direction: Here, the fluorescence signals are consistent with single atom dynamics, making the amplitude of the first peak of the Rabi oscillation independent of $N$; this is highlighted by the white filled diamonds in Fig.\,\ref{fig2}(a). 

To understand the observed behaviors, we start from the scaled rate of photon emission in a direction $\boldsymbol{\hat{k}}$, the observable measured by the APDs, given by \cite{allen1987optical}
\begin{equation}
\bar\gamma(t,\boldsymbol{k})=\frac1N\sum_n\left[\langle\hat{e}_n\rangle (t)+\sum_{m\neq n}
e^{i \boldsymbol{k}\cdot(\boldsymbol R_m-\boldsymbol R_n)}\langle\hat{\sigma}^+_m\hat{\sigma}^-_n\rangle (t)\right]
\label{eq1}
\end{equation}
where $\boldsymbol{k}\equiv 2\pi /\lambda_0 \, \boldsymbol{\hat{k}}$, $\boldsymbol R_n$ the position of the $n$-th atom with internal states $\ket{g_n},\ket{e_n}$ and operators $\hat{e}_n \equiv \ket{e_n}\bra{e_n}$ and $\hat{\sigma}^-_n \equiv\ket{g_n}\bra{e_n}=(\hat{\sigma}_n^+)^{\dagger}$. Superradiance originates from the second term in Eq.\,(\ref{eq1}) describing  the correlations between the atoms. 
In the case of independent atoms, the light emitted by the cloud is proportional to the population inversion of each atom $\langle\hat{e}_n\rangle (t)$ [first term in Eq.~\eqref{eq1}]. In the axial direction, however, the values of the peak over steady-state ratio shown in Fig.~\ref{fig2}(a) cannot be explained without the second term in Eq.\,(\ref{eq1}). This indicates the presence of phase correlations along the main axis of the cloud [second term in Eq.~\eqref{eq1}]. Importantly, this phase coherence is not imposed by the driving laser since the direction of superradiance is perpendicular to it: In a state created by the laser drive (neglecting spontaneous emission): $\ket{\psi_{\rm las}}=\Pi_{\otimes n}(\cos\theta\ket{g_n}+e^{i\boldsymbol k_{\rm las}\cdot\boldsymbol R_n}\sin\theta\ket{e_n})$, the second term [$\propto\sum_n\sum_{m\neq n}e^{i (\boldsymbol k_x-\boldsymbol k_{\rm las})\cdot(\boldsymbol R_m-\boldsymbol R_n)}$] averages to 0. The phase relation responsible for superradiance thus emerges during emission, and is imposed by the cloud geometry. More precisely, the Fresnel number for our geometry is $F=\pi \ell_{rad}^2/\lambda_0 \ell_{ax} \simeq0.05 \ll 1$, and, due to diffraction, the axial spontaneous emission involves all atoms of the cloud (in a single spatial mode) \cite{gross1982}. This is in analogy with cQED, where the external cavity induces a preferential emission mode. In the radial direction, contrastingly, this condition is not fulfilled as $F\gg1$: spontaneous emission is not collective, and the recorded temporal traces are $\propto \sum_{n} \braket{\hat{e}_{n}}$.
\par
We report in Fig.\,\ref{fig2}(b) the experimental measurements of the Rabi frequency, $\Omega$, as a function of $N$. This quantity is determined by Fourier-transforming the temporal traces of the Rabi oscillations, and fitting the resulting spectra with a Gaussian distribution. The extracted frequencies are compared to that of a single atom, i.e.., $\Omega/\Gamma_0=\sqrt{s/2}$.
We observe that the Rabi frequency of the system is independent of $N$, despite the enhancement of light emission in the axial direction. This indicates that the ensemble's coupling to the driving laser is not modified by superradiance. In our situation, superradiance alters only spontaneous emission. Furthermore we observe that not only do the heights of the photon emission peaks increase with $N$, so does their temporal position [inset of Fig.\,\ref{fig2}(b)]. This suggests that superradiant correlations take some time to emerge. The fact that we observe unchanged Rabi oscillations in the radial direction indicates that, in our regime, superradiance very weakly modifies the population dynamics. This in turn suggests a hierarchy of timescales where the Rabi period is shorter than the typical superradiance time ($\tau_{\rm S}$): $\Omega\gtrsim\tau_{\rm S}^{-1}>\Gamma_0$. 


As indicated by Eq.~\eqref{eq1}, a theoretical prediction of the observed emission dynamics requires calculating two-operator correlations. These can be calculated from the density matrix, whose time evolution is governed by a master equation, which includes dipole-dipole couplings between the atoms \cite{Lehmberg1970Radiation,Agarwal1970Master,gross1982, CARMICHAEL2000417}. Despite the knowledge of the microscopic details of our ensembles, the application of exact numerics is not feasible due to the large number of atoms.
We thus make use of an approximate treatment based on a truncation of the operator equations as described in Ref.~\cite{robicheaux2021lowest, plankensteiner2021quantumcumulantsjl}.  
Briefly, the equations for the expectation value of products of $n$ operators depend on the expectation value of products of $n+1$ operators, etc. By using cumulants to approximate contributions of higher order terms the hierarchy can be truncated, and the equations can be closed to a given order.
For example, the second-order mean-field approximation (MF2) replaces three operator expectation values 
with products of one and two operator expectation values assuming the cumulants for the three operators are zero \cite{kubo1962generalized}, e.g., $\langle \hat{e}_l\hat{\sigma}^-_{m}\hat{\sigma}^+_{n} \rangle \to \langle \hat{e}_l \hat{\sigma}^-_{m}\rangle \langle \hat{\sigma}^+_{n} \rangle + \langle \hat{e}_l \hat{\sigma}^+_{m}\rangle \langle \hat{\sigma}^-_{n} \rangle + \langle \hat{\sigma}^-_m \hat{\sigma}^+_{n}\rangle \langle \hat{e}_l \rangle-2\langle \hat{e}_l \rangle\langle \hat{\sigma}^-_{m} \rangle\langle \hat{\sigma}^+_{n} \rangle$. In contrast to the early approach to superradiance, described for example in \cite{allen1987optical, rehler1971super}, this approximation accounts for dipole-dipole interactions between emitters and does not impose any a priori coherence in the many-body wavefunction. These simulations can also include an external drive. The differential equations for the operators were solved numerically for fixed positions of the atoms. 
Different random configurations were averaged until a total of $\sim 20,000$ atoms was reached. Because even the MF2 approximation is computationally intensive, most calculations were done with 20, 40, 80, ... 5120 atoms and compared with the closest experimental number. The positions were chosen randomly using a thermal distribution that matches the size of the atomic cloud.  
Because the CPU and memory requirements increase dramatically going from MF2 to the next order, where the cumulant of the four-body operator is set to zero, i.e..~MF3, we were not able to establish the errors resulting from the MF2 approximation for the experimental parameters. We did, however, perform calculations with 10, 20, 40, and 80 atoms at the MF2 and MF3 level for larger densities where the collective emission rate deviates from the single atom results by more than a factor of two. In these conditions, the MF2 and MF3 calculations of $\bar\gamma(t,\boldsymbol{k})$, differ by less than $\sim 5\%$.

The  results of our simulations are reported in Fig.\,\ref{fig2}. They   reproduce the trend observed in the experimental data, but only qualitatively. The mismatch might be due to a concatenation of various factors that individually would be negligible. These include: 
a non-perfect knowledge of the density distribution of the cloud, depumping effects, effects of the atomic motion, atomic losses during the excitation protocol, and fluctuations in the intensity of the driving field. Despite this, the agreement between experimental and numerical results is remarkable, since the theoretical model does not use any free parameters to fit the data. Importantly, a mean-field approach \cite{do2020collective, Glicenstein2020,Bettles2020}, where $\langle\hat{\sigma}^+_m\hat{\sigma}^-_n\rangle  \to \langle\hat{\sigma}^+_m \rangle \langle \hat{\sigma}^-_n\rangle$ (MF1), is unable to reproduce the data, 
even qualitatively; the results we obtain for the mean field approximation are always consistent with single-atom OBEs for the experimental parameters. This highlights the crucial role of two-atom correlations in our observations, which are not accounted-for in the mean-field treatment but are captured by the MF2 model.

The numerical simulations allow the evaluation of the total photon emission rate per atom using Eq.\,(32) of Ref.\,\cite{robicheaux2021lowest}, 
reported in the inset of Fig.\,\ref{fig2}(a). In the large $N$ regime $\bar\Gamma(t)$ is found to be larger than in the small $N$ case, confirming that the enhanced emission in the axial direction is not due to a reduction in other directions, but to an enhanced scattering rate.
This enhancement could help bring superradiant lasers to power levels suitable for practical applications \cite{meiser2009prospects}.

\begin{figure}
	\centerline{\includegraphics[width=0.95\columnwidth]{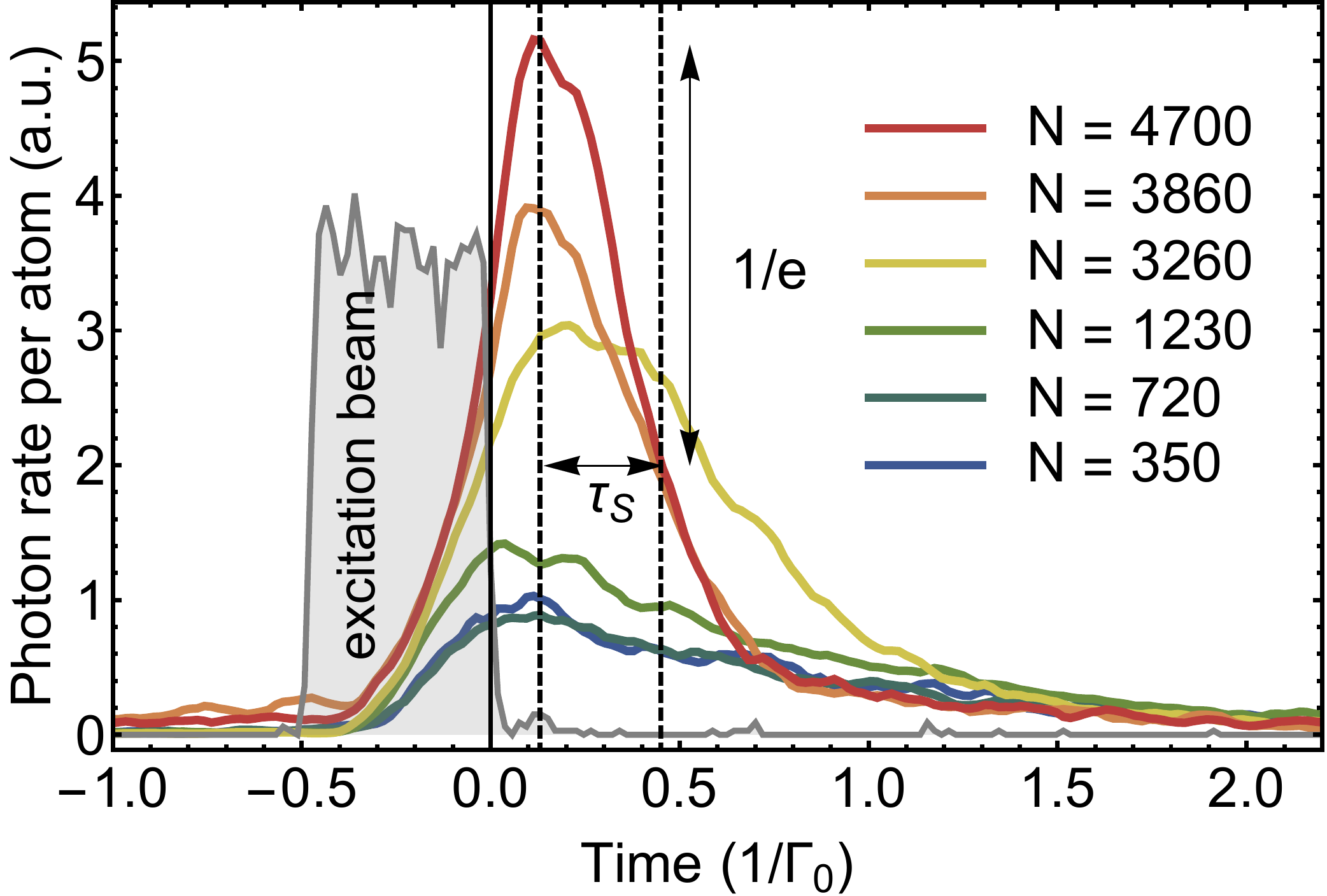}}
	\caption{{\bf Observation of superradiant emission for an inverted system}. 
	Examples of experimental photon rates recorded along the axial direction of the cloud, normalized by the value of $N$ for the cloud. The definition of the characteristic superradiant time (see text) is schematically shown on the $N=4700$ trace. The black vertical line represents the end of the excitation pulse, located at $t=0$, the gray shaded area shows a measured pulse for reference.}
	\label{fig3}
\end{figure}

The observation of the collective Rabi oscillations reported above shows that superradiance does
take place in our driven atomic cloud, but that the resonant drive is strong enough to impose 
a population inversion. This opens the way to the direct investigation of Dicke superradiance, 
i.e..~the collective decay of an inverted system after switching off the driving field. 
Our system allows us to study two-level atoms 
in an ensemble with dimensions close to the transition wavelength, approaching the idealized model that Dicke originally proposed \cite{dicke1954}. 
Moreover, as the atomic cloud is dense, the influence of the light-induced, 
dipole-dipole interactions between atoms can be investigated. 

\begin{figure}
	\centerline{\includegraphics[width=1\columnwidth]{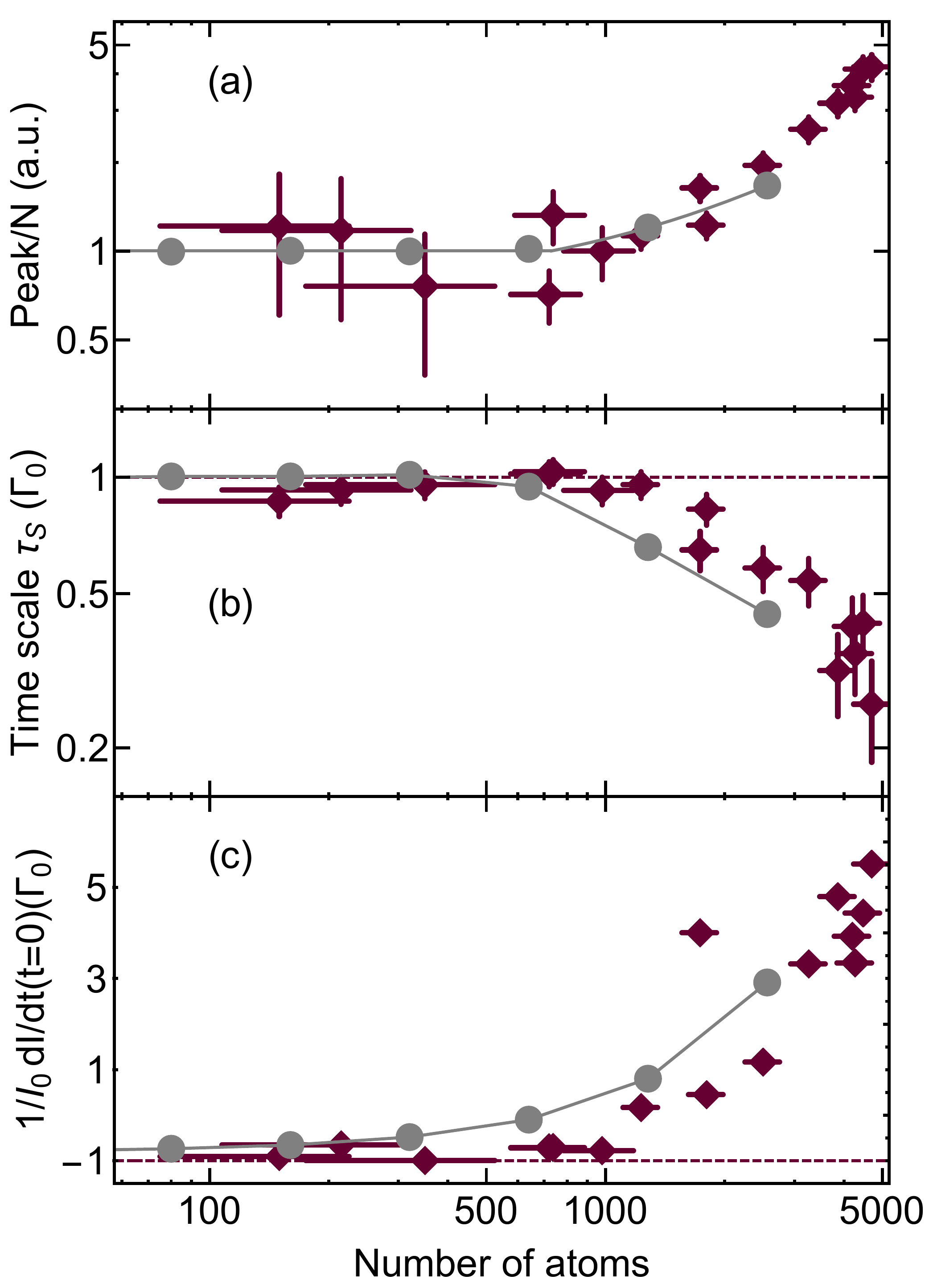}}
	\caption{{\bf Analysis of the superradiant decay}
	(a) Peak photon emission in the axial direction of the superradiant burst normalized by $N$ as a function of $N$. The error bars are the quadratic sum of the standard error on the peak position and on $N$, which is the largest contribution. 
	(b) and (c): $1/e$-decay time and initial slope of the superradiant 
	emission rate after switching-off the driving laser ($t=0$). In (b) the error bars represent the temporal resolution of the detector while in (c) they are evaluated from the errors in the linear fit performed at the end of the pulse. 
	Gray circles: results of the numerical simulations using the MF2 model.
	The dashed lines in (b) and (c) represent the behavior in the single atom case.}
	\label{fig4}
\end{figure}

We report examples of experimental traces acquired \textit{along the axial direction} of the cloud for different $N$ in Fig.\,\ref{fig3}. We observe that, as $N$ increases, the photon emission switches from an exponential decay to a short burst.  However, since the duration of the $\pi$-pulse is comparable to the time-scale of the enhanced decay rate, superradiant emission should start before the end of the excitation pulse. This is what we observe in 
Fig.\,\ref{fig3}: the intensity emitted {\it per atom} 
at the end of the pulse increases with $N$, 
while, ideally, it would be independent of $N$ \cite{dicke1954}. Despite this, 
we observe that, as $N$ increases, the emission maximum of the cloud increases \textit{after} the drive is switched off. 
Additionally, as highlighted by the temporal narrowing of the burst, 
the timescale characterizing the collective decay decreases as $N$ increases.

In order to quantitatively investigate these features, we report in Fig.~\ref{fig4}(a) the measured peak intensity per atoms as a function of $N$. It displays a plateau for $N\lesssim 1350$ before increasing linearly above this threshold. This trend shows that, along the long axis of the cloud, the intensity of the light emitted scales as $N^2$ for large $N$. This scaling, as well as the existence of a threshold in $N$, are typical fingerprints of Dicke superradiance. In our cloud, the existence of a threshold is due to the fact that the axial size is larger than the wavelength, necessitating larger values of $N$ to compensate \cite{gross1982,allen1987optical}. In Fig.\,\ref{fig4}(b), we report the measured timescale of the superradiant burst, defined as the time difference between the intensity maximum and the time at which the fluorescence emitted by the cloud decays to $1/e$ of its maximum value (see Fig. \ref{fig3}) \footnote{We use this definition since, for large $N$, the temporal shape of the burst is not an exponential. For small $N$, where the fluorescence does decay exponentially, our definition coincides with the usual decay time.}. Finally, a linear fit in a 5~ns-temporal window centered around $t=0$, yields the emission rate, i.e.., the initial slope of the decay at the switch-off of the driving reported in Fig.\,\ref{fig4}(c). We perform MF2 calculations also for this experiment, studying the dynamics of a system where the atoms are prepared in the  state $\ket{\psi_{\rm las}}$ written above, with $\sin^2\theta=0.9$, i.e.~$90\%$ in the excited state. The results, reported in Fig.\,\ref{fig4}, agree quantitatively with the data. This agreement indicates that despite superradiance occurring during the driving, our system approximately reproduces Dicke's scenario. This observation also demonstrates that resonant dipole-dipole interactions do not prevent the onset of superradiance at our densities, and should not hinder the performances of superradiant lasers if the density is increased to improve laser power beyond the densities currently used. At much higher densities, a suppression of superradiance is expected in disordered clouds \cite{superradiance2017sutherland}, as opposed to ordered arrays \cite{masson2020,masson2021universality}.



In conclusion, we have observed supperradiance in a disordered cloud of two-level atoms. This superradiance emerges from a strong coupling of the atomic cloud with a single mode, thus realising ``cavity-less cQED" experiments. We established that despite a resonant drive perpendicular to the superradiant mode propagation direction, correlations do emerge leading to superradiance. The direction of superradiant emission is thus set by the geometry of the cloud rather than by the driving laser direction, as opposed to what is typically assumed \cite{allen1987optical}. In this situation superradiance is predicted theoretically only when accounting for two-atom correlations. 
Finally, there are other manifestations of superradiance that could be investigated. As an example, an interesting future direction of research would be the study of intensity correlations of the emitted field, which might exhibit two-photon correlations impacted by superradiance and resonant dipole-dipole interactions.  
\begin{acknowledgements}
{We thank Martin Robert de Saint Vincent and Bruno Laburthe for discussions, and L.~Ulrich, S.~Welinsky and P.~Berger for the loan of a fiber EOM for preliminary tests. This project has received funding from the European Union’s Horizon 2020 research and innovation program under Grant Agreement No. 817482 (PASQuanS), Agence Nationale de la Recherche (project DEAR) and by the Région Île-de-France in the framework of DIM SIRTEQ (projects DSHAPE and FSTOL). A.\,G.~is supported by the Délégation Générale de l'Armement Fellowship No. 2018.60.0027.
F.\,R.~is supported by the National Science Foundation under Award No. 1804026-PHY.}
\end{acknowledgements}

\bibliography{biblio}

\setcounter{figure}{0}
\renewcommand\thefigure{A\arabic{figure}} 
\setcounter{equation}{0}
\renewcommand\theequation{A\arabic{equation}} 
\appendix

\end{document}